\documentclass[usenatbib]{aastex}

\usepackage{graphicx}
\usepackage{epsfig}
\usepackage{amsmath,amssymb}
\usepackage{verbatim}
\usepackage{epstopdf}
\usepackage{pdflscape}
\usepackage{subfigure}
\usepackage{chngcntr}
\usepackage{float}
\usepackage{placeins}
\usepackage{color}
\usepackage{gensymb}
\usepackage{amsmath}
\usepackage{tabularx}
\usepackage{array}

\def\lapprox{\lower.4ex\hbox{$\;\buildrel <\over{\scriptstyle\sim}\;$}}
\def\gapprox{\lower.4ex\hbox{$\;\buildrel >\over{\scriptstyle\sim}\;$}}

\begin{document}

\shorttitle{Bent Lobes for DLRGs }
\shortauthors{Silverstein, Anderson, and Bregman}
\title{Increased Prevalence of Bent Lobes for Double-Lobed Radio Galaxies in Dense Environments  }
\author{Ezekiel M Silverstein\altaffilmark{1}, Michael E Anderson\altaffilmark{2}, and Joel N Bregman\altaffilmark{1}}

\altaffiltext{1}{Department of Astronomy, University of Michigan, Ann Arbor, MI 48109; \hspace{3 cm} ezmasilv@umich.edu, jbregman@umich.edu.}
\altaffiltext{2}{Max-Planck Institute for Astrophysics, Garching bei Muenchen, Germany; michevan@mpa-garching.mpg.de}

\begin{abstract}
	Double-lobed radio galaxies (DLRGs) often have radio lobes which subtend an angle of less than 180 degrees, and these bent DLRGs have been shown to associate preferentially with galaxy clusters and groups. In this study, we utilize a catalog of DLRGs in SDSS quasars with radio lobes visible in VLA FIRST 20 cm radio data. We cross-match this catalog against three catalogs of galaxies over the redshift range $0 < z < 0.70$, obtaining 81 tentative matches. We visually examine each match and apply a number of selection criteria, eventually obtaining a sample of 44 securely detected DLRGs which are paired to a nearby massive galaxy, galaxy group, or galaxy cluster. Most of the DLRGs identified in this manner are not central galaxies in the systems to which they are matched. Using this sample, we quantify the projected density of these matches as a function of projected separation from the central galaxy, finding a very steep decrease in matches as the impact parameter increases (for $\Sigma \propto b^{-m}$ we find $m = 2.5^{+0.4}_{-0.3}$) out to $b \sim 2$ Mpc. In addition, we show that the fraction of DLRGs with bent lobes also decreases with radius, so that if we exclude DLRGs associated with the central galaxy in the system the bent fraction is 78\% within 1 Mpc and 56\% within 2 Mpc, compared to just 29\% in the field; these differences are significant at $3.6\sigma$  and $2.8\sigma$ respectively. This behavior is consistent with ram pressure being the mechanism that causes the lobes to bend. 
\end{abstract}

\keywords{galaxies: clusters: intracluster medium, intergalactic medium, galaxies: jets, quasars: general }

\maketitle

\section{Introduction}

	Double-lobed radio galaxies (DLRGs) are spectacular sights in the radio sky, and also are scientifically interesting because they connect processes on the $\sim$AU scale of a galaxy's supermassive black hole (SMBH) to the extragalactic scale ($\sim$tens-hundreds of kpc).  Such galaxies are historically divided into two classes \citep{Fanaroff1974}:  less-luminous FR I galaxies with brighter cores and fainter lobes, and  more-luminous FR II galaxies with brighter lobes and fainter cores.   FR I galaxies also tend to be found in optically luminous galaxies \citep{Ledlow1996}, typically in brightest cluster galaxies (BCGs), the most luminous galaxies of all \citep{Zirbel1997}.  FR II galaxies are also found in denser environments, but preferentially in groups rather than clusters \citep{Zirbel1997}.

	Bent-double radio galaxies are a subclass of DLRGs, with the angle between their two lobes bent so that they subtend less than $180\degree$.  They are more likely to be found in high-density environments than ordinary DLRGs, and are found with roughly equal probability in clusters and groups; in total, 6\% of Abell clusters host a bent-double galaxy \citep{Blanton2001}. \citet{Wing2011} explore the use of bent DLRGs as a way to detect galaxy clusters, and are able to associate 78\% of their sample of bent DLRGs with clusters or rich groups in SDSS. This correlation with environment may or may not be causal, but there are several plausible mechanisms which may explain it.
	
	One such mechanism is ram pressure experienced by the lobes as the galaxy moves through the intragroup/intracluster medium \citep{Miley1972,Jaffe1973,Jones1979}.  There are also other possibilities, such as collisions between outflowing lobes and other cluster galaxies \citep{Stocke1985} or merger-induced precession of the SMBH spin axis \citep{Merritt2002}. Ongoing or recent mergers \citep{Roettiger1996}, or clusters with sloshing motions in their intracluster medium (e.g. Abell 2029; \citealt{Paterno2013}) could also be important \citep{Mendygral2012}.  Each mechanism predicts that more bent DLRGs should arise in dense environments, but there are perhaps second-order observable differences which may be used to distinguish between them (e.g. \citealt{Rector1995}).

	Regardless of which mechanism drives the relationship between bent doubles and environment, a few researchers have begun to invert the relationship, using bent doubles to probe the diffuse gas around the galaxies.  \citet{Freeland2008} examined two FR I bent DLRGs, one inside a small group and the other at a projected distance of 2 Mpc from a group, and inferred intergalactic gas densities of $4 \times 10^{-3}$ cm$^{-3}$ and $9 \times 10^{-4}$ cm$^{-3}$, respectively, at the locations of these galaxies (assuming the bending was caused by interaction with diffuse intergalactic gas).  \citet{Freeland2011} expanded this analysis to seven bent DLRGs, and found similar results.  \citet{Edwards2010} discovered a bent-double radio galaxy at a projected distance of 3.4 Mpc from the center of Abell 1763, from which they inferred the presence of an intercluster filament and provided loose constraints on its density. Finally, \citet{McBride2014} study the bending of the lobes of NGC 1272 induced by the Perseus cluster and are able to constrain the pressure in the lobes. 

	Here we extend this type of analysis to a much larger sample of bent DLRGs, using the catalog of DLRGs compiled by \citet{deVries2006}.  We cross-match this sample of DLRGs with various catalogs of central galaxies massive halos, which collectively span a significant fraction of Cosmic time.  With this dataset, we can study the environmental behavior of DLRGs in unprecedented detail.

	The structure of this paper is as follows.  In Section 2, we discuss the catalogs examined in this work, the various selection criteria which were used to generate them, and the methods for cross-matching the catalogs. In Section 3 we analyze the results of this cross-matching in order to measure the environmental behavior of DLRGs and the properties of the bending.  In Section 4, we interpret these results and conclude.


\section{Sample and Methods}

We consider in this paper the catalog of DLRGs from De Vries, Becker, and White (2006; hereafter DBW). They cross-matched 44894 quasars from the Sloan Digital Sky Survey (SDSS) Data Release 3 with the Faint Images of the Radio Sky at Twenty centimeters (FIRST; \citealt{Becker1994}) survey from the Very Large Array (VLA) in order to construct a very large sample of DLRGs.  For each SDSS quasar, DBW examined each radio source projected within 450$^{\prime \prime}$, using a pairwise ranking system in order to evaluate the probability of the radio sources being lobes of the central quasar.  Their ranking system favors potential sources which are closer in the sky to the central quasar and which have larger opening angles.

	From the DR3 sample of 44894 SDSS quasars, DBW identified 35936 candidate DLRGs.  A significant fraction of these candidate DLRGs are ``false positives'' - quasars for which two radio sources are projected by chance in the sky such that the algorithm of DBW identifies them as potential radio lobes.  DBW studied the incidence of these ``false positives'' and found that, for pairs of radio sources around a quasar with a projected separation of less than 90'', a large majority of the candidate DLRGs are real DLRGs (especially for opening angles close to $180 \degree$).  Candidate DLRGs with projected separations of 60''-120'' are about equally likely to be real DLRGs or false positives.  Based on these results as well as our own studies of these populations, we select the 780 DLRG candidates with projected lobe separations less than 90'' for further study.  The remaining 780 candidate DLRGs have redshifts ranging from $z=0.041$ to $z=4.889$, and there is no single catalog tracing large-scale structure in SDSS over such a wide range of redshifts.  We therefore created a composite sample using three different catalogs spanning different redshifts.

\subsection{Groups and Clusters}

	The first bin of galaxy groups and clusters spans $z=0$ to $z=0.20$.  This entire volume is covered by a flux-complete (down to Galactic-extinction-corrected Petrosian $r$-magnitude of 17.77) group and cluster catalog \citep{Tempel2014} containing 82458 groups and clusters.
 
	At higher redshift it is more difficult to identify groups and clusters using the relatively shallow SDSS photometry.  Instead, we use the catalogs from the SDSS-Baryon Oscillation Spectroscopic Survey (BOSS) Data Release 10 \citep{Ahn2014}.  BOSS is a spectroscopic survey of massive galaxies in the SDSS footprint.  There are two sets of catalogs - LOWZ and CMASS - with slightly different photometric selection criteria.  The selection criteria are designed such that both catalogs are approximately stellar-mass limited with typical log $M_{*}/M_{\odot}$ = 11.3 \citep{Parejko2013, Guo2013}.  At these stellar masses, the BOSS galaxies are predominantly red and highly clustered \citep{Guo2013}, so we take them to be reasonably good tracers of large scale structure at these redshifts.  We therefore select galaxies from the LOWZ catalog with $0.20<z<0.47$ for our second redshift bin, and galaxies from the CMASS catalog with $0.47<z<0.70$ for our third redshift bin.  We placed cuts at $z=0.2$, $z=0.47$ and $z=0.7$ to ensure there was not any double-counting between different catalogs.  These bins contain 209788 and 446158 central galaxies, respectively.
	
Together, our total list of galaxies, groups, and clusters contains 738404 systems. While these systems are primarily galaxy groups and clusters, our coordinates for the BOSS sample will refer to the bright central galaxies, and we will often refer to the galaxies for simplicity, although of course the larger group/cluster is the primary object of interest.

\subsection{Cross Matching}

	We cross-match the DBW catalog of DLRGs with the galaxies in our various redshift bins.  Our initial match criterion is a DLRG falling within 10 projected Mpc and 3,000 km/s of the galaxy. We compute the projected distance (impact parameter) from the measured angular separation, which we multiply by the comoving distance of the galaxy estimated from its redshift assuming the \citet{Planck2015} cosmological parameters. The radial velocity separation cutoff of 3,000 km/s was chosen as a somewhat arbitrary upper limit on the escape velocity of a large cluster. We find that there is a steep drop-off of DLRG-group pairs at velocity separations greater than 3,000 km/s, so our results are not very sensitive to the exact cutoff employed here.
	
Using this cross matching criterion, we found 81 DLRG - galaxy pairs.  This method of cross-matching allows for the possibility of a DLRG matching with multiple galaxies, which happens in most cases (61/81).  We therefore incorporate the galaxies' impact parameters, velocity separations and halo masses to estimate a relative probability for the DLRG to be associated with each matched galaxy. Under the simplest assumption of that the galaxies in a group are isotropically distributed as $R^{-3}$ in 3D space (which is the NFW scaling at the Mpc scales we consider here) and that they have an isotropic and Gaussian distribution in velocity space, their projected space density will scale with impact parameter $b$ as $(b/b_0)^{-2}$ and their projected velocity density will scale with velocity separation $\sigma$ as $e^{-\sigma^2/2\sigma_0^2}$. We therefore define the relative probability for a DLRG to be associated with a system $i$ as
	
	\begin{equation} P_{i} = C \times (M_{i} / \sigma_{0}) \times (b/b_{0})^{-2} \times e^{-\sigma^2/2\sigma_{0}^{2}} \end{equation}
	
\noindent 

In this expression, $M_i$ is the mass of the cluster or group, and absorbs the mass dependence of $b_0$ and $\sigma_0$, allowing our estimator to prefer to associate DLRGs with more massive systems. $C$ is a normalization constant defined for each DLRG such that $\Sigma_i P_i = 1$, and the $\sigma_0$ appears in the denominator in order to normalize the Gaussian term to unity.  We set $b_0 = 300$ kpc and $\sigma_0 = 1100$ km/s, which are typical values for a massive cluster, although we varied these values by $\pm25\%$ and the matches were not significantly changed. 

	For the groups with redshifts $z<0.2$, we define $M_i$ from the mass estimations  included in the Tempel et al. (2014) catalog, which are based on the measured velocity dispersion of the galaxies in the group and an assumed halo mass profile (Navarro et al 1997; Maccio et al. 2008).  However, not many galaxies are detected in most of the groups --  60\% have two galaxies, and  91\% have five or fewer galaxies -- so these mass estimates are quite uncertain.  For the groups with five or fewer detected galaxies, 71\% have masses less than $1\times10^{13}M_{\odot}$ in the Tempel et al. catalog, but for such poor groups the uncertainties in measuring velocity dispersion become more important than the measurement itself, and for simplicity  we institute a mass floor of  $1\times10^{13}M_{\odot}$ for these poor groups.

For the catalog containing groups with redshifts $0.2<z<0.7$, we have no straightforward halo mass estimator, so we assign each system the same halo mass (the exact value drops out of equation 1, but we use $5\times10^{13}M_{\odot}$; \citealt{vanUitert2015}).

For each DLRG, the galaxy with the largest $P_i$ is selected as the match. Of the 61 DLRGs with multiple potential matches, 33 are matched to the galaxy with the lowest impact parameter out of the potential candidates. In angular space, the median impact parameter is 7.7', and in physical space it is 3.3 Mpc.

\subsection{Verification of DLRGs} \label{verification}

We also visually inspected each of the 81 candidates using the VLA-FIRST images. Since the DLRG catalog was assembled automatically, many ``false positives'' are obvious by eye, with one or both lobes missing, and/or the outlying sources clearly appearing as point sources instead of radio lobes. 

	We also performed additional quantitative tests to verify the reality of the DLRGs in our sample. First, we require that the sum of the specific luminosities from the central galaxy and the two lobes be brighter than $1\times10^{31}$ erg s$^{-1}$ Hz$^{-1}$. This is ten times fainter than the specific luminosity separating FR II objects from FR I objects \citet{Fanaroff1974}, and serves as a test that the luminosity of the DLRG is physically plausible. No k-correction is applied for this calculation. Only five of the 81 DLRGs fail this test, and visual inspection shows that all five of these objects appear to be projections of unrelated radio point sources.
	
	Second, we require the lobes to be diffuse objects, not point radio sources, so we discard any DLRG if one or both of its lobes appear less than 2.5'' in radius.  This was done by visually assessing the diameter of each lobe in the VLA FIRST images.  56 of the 81 DLRGs pass this cut while 25 fail.
	
	Third, we require the DLRG to be the brightest radio source within a 1$^{\prime}$ radius region to ensure that the radio lobes are not mistakenly attributed to the central QSO. In a few cases, we disagreed with the choice of central galaxy in the DWB catalog (i.e. the central galaxy was misidentified as a lobe and vice versa; this is obvious upon visual inspection but difficult to quantify algorithmically). In these cases we manually changed the lobe placement if it was clear the original placement was incorrect and there was a clear alternative associated with the DLRG core.  When there was not a clear alternative, we rejected the DLRG.  When a DLRG was accepted yet needed a lobe position update, we repositioned the lobes and recalculated the angle between the DLRG lobes using these new coordinates.  In most cases the change is small, but for a handful of objects we identified one of the lobes with a different radio source than DBW, which caused the bending angle to change significantly.  In 13/81 cases the newly calculated angle changed by at least $10\degree$, but 7 of these 12 cases were rejected by one of the other tests outlined above. The six remaining DLRGs with changed angles have ID numbers of 17, 18, 22, 37, 38, and 41 in Table 1 below.

All objects that fail one or more of these requirements are rejected, along with the objects that failed the visual classification. In the end, we accepted 44 DLRG-group matches from the original 81. Of these 44 DLRGs, ten are matched to a different cluster or group than the one with the smallest impact parameter, due to the $M_i$ and $\sigma$ terms in equation (1). In Table 1 we present basic data for these 44 DLRGs.

\begin{table}[]
\caption{Data for DLRGs and Matched Galaxies}
\scriptsize
\setlength{\extrarowheight}{-.5em}
\begin{tabular}{lllllllll}
N & ra & dec & z & angle & + , - angle err & galaxy ra & galaxy dec & galaxy z \\
\hline
1 & 115.35507 & 33.55558 & 0.364 & 179 & 1, 3 & 115.524 & 33.242 & 0.37118 \\ 
2 &122.13905 & 42.81011 & 0.543 & 174 & 6, 26 & 122.055 & 42.86 & 0.53488 \\ 
3 & 123.32854 & 50.21106 & 0.571 & 174 & 6, 11 & 123.331 & 50.21 & 0.57335 \\ 
4 & 125.39003 & 47.04369 & 0.128 & 151 & 29, 45 & 125.435 & 47.133 & 0.12582 \\ 
5 & 129.66907 & 47.56963 & 0.695 & 173 & 7, 12 & 129.645 & 47.721 & 0.69409 \\ 
6 & 132.66646 & 54.6315 & 0.367 & 174 & 6, 7 & 132.466 & 54.684 & 0.3641 \\ 
7 & 137.08757 & 4.84985 & 0.524 & 174 & 6, 17 & 137.042 & 4.804 & 0.52406 \\ 
8 & 138.50735 &5.13073 &0.301& 176 & 4,7& 138.725& 5.575& 0.30186\\
9 & 140.60475 &43.1304& 0.236& 172 & 8,6 &140.636& 42.784& 0.2281\\
10 & 141.90874 & 1.742 & 0.419 & 180 & 0, 9 & 141.894 & 1.893 & 0.41805 \\ 
11 & 145.26669 & 38.8975 & 0.616 & 177 & 3, 6 & 145.18 & 38.747 & 0.62589 \\ 
12 &146.93811& 7.42238& 0.086 &175 & 5,19 & 146.918& 7.424& 0.08749\\  
13 & 147.86035 & 1.78106 & 0.495 & 179 & 1, 14 & 147.667 & 1.754 & 0.48888 \\ 
14 & 152.25861 & 7.22885 & 0.456 & 160 & 8, 3 & 152.035 & 7.12 & 0.46281 \\ 
15 & 155.27518 & 45.39219 & 0.364 & 177 & 3, 8 & 155.527 & 45.63 & 0.36814 \\ 
16 & 157.9313 & 52.42644 & 0.166 & 179 & 1, 12 & 157.554 & 52.797 & 0.16837 \\ 
17 & 159.67511 & 4.55238 & 0.423 & 180 & 0, 9 & 159.697 & 4.361 & 0.42981 \\ 
18 & 163.7514 & 52.03359 & 0.187 & 141 & 31, 4 & 164.376 & 51.669 & 0.19216 \\ 
19 & 164.22566 & 5.28702 & 0.456 & 156 & 17, 17 & 164.174 & 5.293 & 0.45756 \\ 
20 & 166.82867 & 10.07159 & 0.633 & 180 & 0, 12 & 166.889 & 9.978 & 0.63287 \\ 
21 & 176.29327 & 1.1823 & 0.626 & 175 & 5, 7 & 176.498 & 1.198 & 0.63159 \\ 
22 & 185.04955 & 2.06174 & 0.24 & 161 & 19, 6 & 185.08 & 1.802 & 0.2354 \\ 
23 & 187.64931 & 9.75526 & 0.638 & 176 & 4, 8 & 187.64 & 9.79 & 0.641 \\ 
24 & 189.0188 & 10.58035 & 0.667 & 180 & 0, 6 & 188.914 & 10.562 & 0.66862 \\ 
25 & 193.75201 & 3.67862 & 0.437 & 137 & 30, 24 & 194.058 & 3.565 & 0.4384 \\ 
26 & 195.9978& 3.65893& 0.184& 177 & 3,29& 195.971& 3.53 &0.18591\\
27 & 197.17867 & 2.72409 & 0.504 & 167 & 7, 9 & 197.178 & 2.881 & 0.50315 \\ 
28 & 205.3952 & 53.74548 & 0.141 & 175 & 5, 15 & 205.42 & 53.431 & 0.1404 \\ 
29 &206.4390& 53.54786& 0.135& 158& 22,8&206.441& 53.381& 0.13705\\
30 & 206.5731& 62.34597& 0.116& 154&26,15& 206.669& 62.5& 0.11549\\
31 & 207.72746 & 5.36847 & 0.442 & 175 & 5, 24 & 207.814 & 5.042 & 0.44039 \\ 
32 & 208.2731 & 4.72743 & 0.523 & 176 & 4, 23 & 208.269 & 4.726 & 0.5238 \\ 
33 & 211.32701 & 4.56859 & 0.352 & 179 & 1, 18 & 211.433 & 4.76 & 0.35014 \\ 
34 & 212.37141 & -1.95491 & 0.638 & 130 & 13, 25 & 212.306 & -1.791 & 0.63794 \\ 
35 & 215.64955 & -1.86979 & 0.666 & 173 & 7, 7 & 215.683 & -1.752 & 0.66582 \\ 
36 & 216.5258 & 40.40889 & 0.664 & 168 & 10, 10 & 216.56 & 40.414 & 0.66398 \\ 
37 & 220.76151 & 52.027 & 0.141 & 129 & 33, 20 & 220.761 & 52.05 & 0.14244 \\ 
38 & 224.7473 & 4.27051 & 0.391 & 144 & 18, 32 & 224.747 & 4.258 & 0.39158 \\ 
39 & 225.34152 & 1.73368 & 0.608 & 160 & 11, 10 & 225.213 & 1.594 & 0.60576 \\ 
40 & 228.0656 & 2.05472 & 0.219 & 159 & 20, 22 & 228.111 & 2.021 & 0.22044 \\ 
41 & 239.32916 & 45.37266 & 0.495 & 143 & 37, 15 & 239.297 & 45.38 & 0.49533 \\ 
42 & 249.73557 & 43.58683 & 0.339 & 174 & 6, 16 & 249.732 & 43.581 & 0.33745 \\ 
43 & 251.43622 & 37.92392 & 0.598 & 176 & 4, 7 & 251.69 & 37.979 & 0.5906  \\
44 & 255.89581 & 39.29323 & 0.523 & 171 & 9, 45 & 255.957 & 39.272 & 0.52014 \\ 
\hline
\end{tabular}
\tablecomments{- List of DLRG - galaxy pairs which pass all of our selection criteria.  The first column shows an identification number for the pair.  The next three columns show the right ascension, declination and redshift of the DLRGs.  Columns 5, 6, and 7 show the angle between the lobes of the DLRG and the +/- errors on these from visual inspection.  The final three columns are the right ascension, declination and redshift of the galaxy to which the DLRG is matched. Note that in most cases the DLRG is not the central galaxy of the cross-matched system, but instead is a satellite galaxy or is outside the virial radius.}

\end{table}


\section{Results}

\subsection{Projected Density of Matches}

The cross-matching, the heterogeneous galaxy catalogs, and the various stages of DLRG verification described in section 2.3 all introduce complicated biases in the sample selection. Modeling the sample selection in detail would require an unwieldy set of assumptions, including assumptions about galaxy evolution, halo occupation, evolution of the DLRG spectral shape and luminosity function, as well as the parameters in which we are interested like the connection between DLRGs and large-scale structure. Since this work is primarily observational, and we want to be as parsimonious with assumptions as possible, we instead construct a ``control'' sample with the same sample selection biases, in order to compare to the DLRG sample.

To do this, we generate a set of mock coordinates for each DLRG by shifting the true coordinates in right ascension and declination by various amount ranging from 4-16 degrees, yielding a total of 28 mock DLRGs for each of the 44 true DLRGs, for a total of 1232 mock DLRGs. We perform the same cross-matching as in section 2.2 for these mock DLRGs. The result is a sample of central galaxies cross-matched with random positions on the sky, but obeying the same redshift distribution as our DLRG sample. 

In Figure 1, we plot the surface density of our DLRG matches as a function of projected radius from the central galaxy. Within each annular bin, we compute the uncertainty from the number of counts, assuming Poisson statistics. Due to the low number of DLRGs, we oversample the surface density  (i.e. bins are spaced by 0.25 Mpc but span 1.0 Mpc in width). This means that goodness of fit tests like the $\chi^2$ statistic are inappropriate (since they rely on the data points being independent of one another), but the error bars on each individual point are unaffected by the oversampling. 

The control sample shows a gradual decrease in projected density as a function of impact parameter, declining by a factor of $\sim 4$ over the 5 Mpc range in impact parameters covered in the plot. This decline is probably due to the algorithm (eq. 1) which favors matches with smaller impact parameters, in combination with our choice to assign the redshifts of the observed DLRGs to the random positions. From 3 Mpc $\lapprox$ b $\lapprox 5$ Mpc, the data and the random sample match very well, implying that chance projections on the sky are a plausible explanation for DLRG - galaxy group/cluster pairs with these large impact parameters.

For $b \lapprox 2$ Mpc, there is a clear increase of these pairs in our data relative to the control sample of chance projections. DLRGs are therefore significantly more likely than random positions on the sky to be found near galaxy groups and clusters. In the innermost bin, the projected density of DLRG - central galaxy pairs is ten times higher than the projected density of random position - central galaxy pairs, and based on Poisson statistics the probability of this occurring by chance is just $3\times10^{-10}$. There is also a small ''dip'' in the observed DLRG projected density at $b\approx2.5$ Mpc, but it is not statistically significant.

Our interpretation of Figure 1 is therefore that the majority of the DLRG - galaxy group/cluster matches reflect physical associations when the impact parameter is less than about 2 Mpc. For the pairs with impact parameter greater than about 2 Mpc the majority (possibly all) of these DLRG - galaxy pairs reflect chance superpositions on the sky. We speculate that there are additional groups which are missing from our catalogs and lie at much smaller projected distances from these DLRGs.

We have also fit a power-law fit to the data in Figure 1 by minimizing $\chi^2$ (which we mentioned is inappropriate for goodness of fit testing due to the oversampling, but still adequate for line fitting). This line has a slope of $m = 1.9\pm0.2$ (with $\Sigma \propto b^{-m}$), corresponding to a real-space decrease in density of $\rho \propto r^{2.9\pm0.2}$. However, this slope is likely too shallow, since the data include a contribution from the "background" of chance superposition as well. We estimate this background by taking the mean from $b=3$ Mpc to $b=4$ Mpc. Subtracting this "background", the slope steepens to $m =2.5^{+0.4}_{-0.3}$. Both of these values imply a very steep decrease with density (i.e. at least as steep as an NFW profile), and this will be discussed further below.

\begin{figure}
\begin{center}
{\includegraphics[width = 7in]{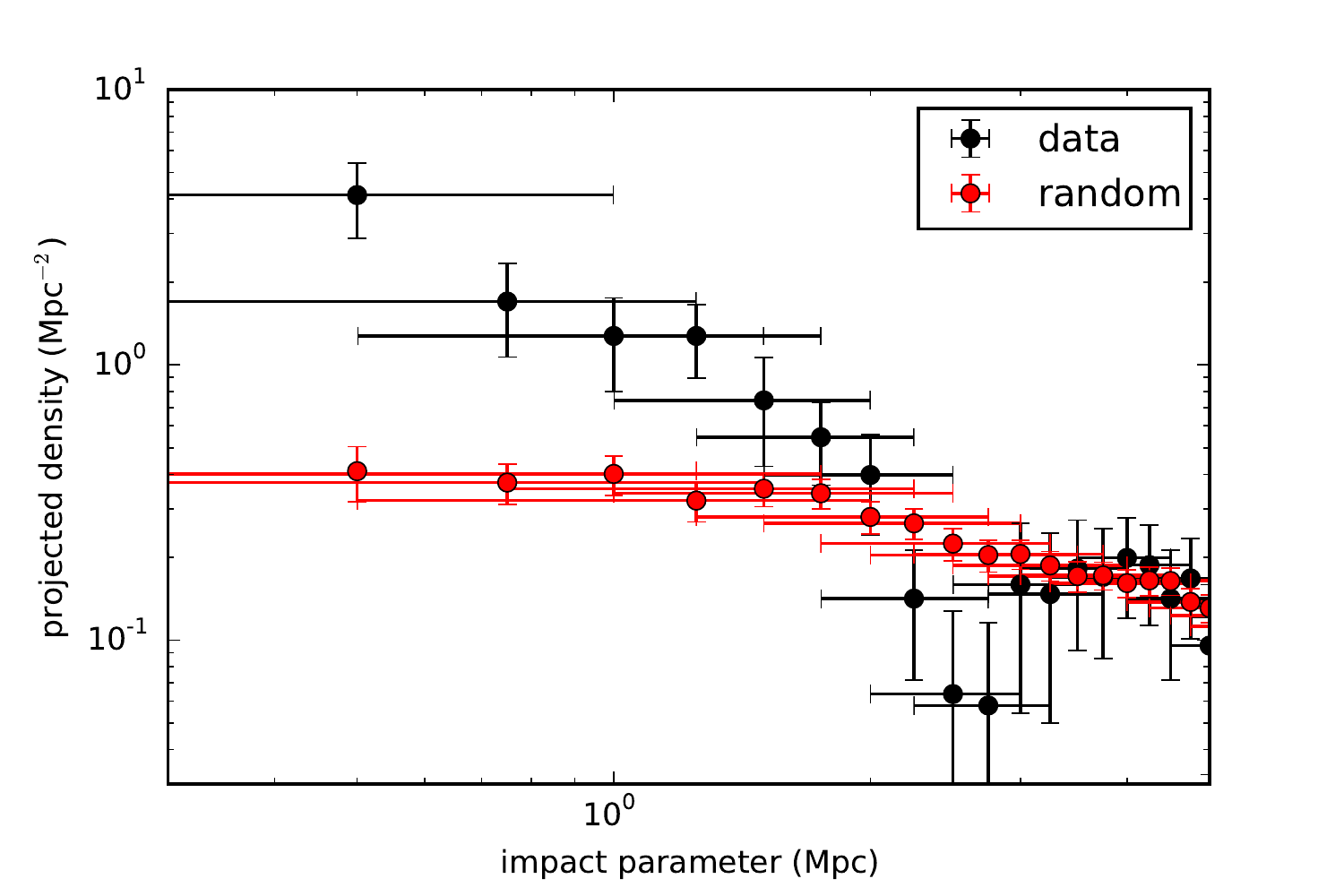}}
\end{center}
\caption{Projected density of DLRG-central galaxy matches (black) and a control sample of random positions - central galaxy matches (red) from our cross-matching analysis. The two populations have roughly the same behavior for impact parameters larger than about 2 Mpc (with the exception of a "dip" at around 2.5 Mpc which is not statistically significant), suggesting that all the DLRG-galaxy matches beyond this radius can be explained as random projections on the sky. For smaller impact parameters, the projected number of DLRGs matching with the central galaxy is much higher than the density of random projections that match, suggesting there is a physical correlation between DLRGs and the galaxies which trace galaxy groups and clusters in our sample. We fit the projected density within 2 Mpc with a power-law, and find a best-fit slope of $m = 1.9\pm0.2$ (with $\Sigma \propto b^{-m}$). We also perform a subsequent fit after subtracting the "background" at large radii and find an even steeper slope: $m = 2.5^{+0.4}_{-0.3}$.}  
\end{figure}

\subsection{DLRG Bending Angles}

In Figure 2 we present the measured angle between the lobes of each DLRG for the sample of 44 verified cross matches. For each of these 44 objects, we have drawn vertical error bars which encompass our uncertainty in the bending angle, as determined by the visual analysis in section 2.3. These are obtained by identifying edges for both lobes and computing all the possible angles which can subtend these edges; these error bars are therefore much more conservative than 1$\sigma$ error bars. 

\begin{figure}
\begin{center}
{\includegraphics[width = 7.0in]{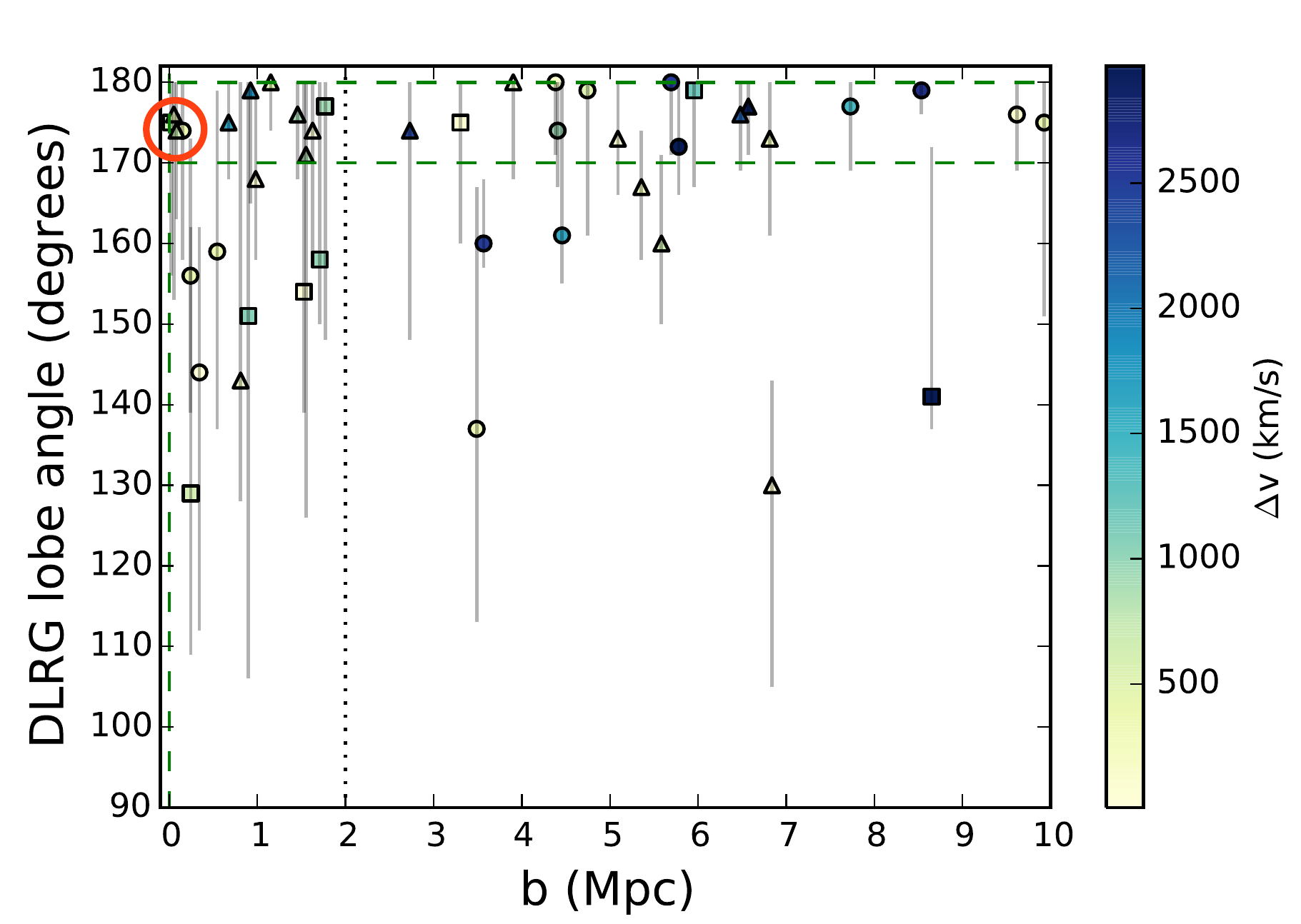}}
\end{center}
\vspace{-1 cm}
\caption{The angle subtended by the DLRG radio lobes plotted against the impact parameter of the DLRG relative to the associated central galaxy.  The horizontal dotted line at $170\degree$ is a cutoff between ``bent'' and ``unbent'' DLRGs; we consider points above this line to be ``unbent'', and draw vertical errors for each point which encompass the full (not 1$\sigma$) range of possible angles for the lobes based on visual inspection.  The colorbar on the right-hand side is the magnitude of the radial velocity difference between the DLRG and the central galaxy.  The red circle in the top left identifies DLRGs we which we argue are the central galaxies of their respective groups. The dotted black vertical line indicated the approximate impact parameter (2 Mpc) within which we see a much higher fraction of bent lobes. DLRG-galaxy pairs at larger impact parameters are likely chance projections on the sky; the DLRG is presumably associated with a closer galaxy group which falls below our detection limit.  Points are given shapes corresponding to their redshift. Circles have $z < 0.2$, squares have $0.2 < z < 0.47$, and triangles have $0.47 < z < 0.7$. }
\end{figure}

We have also drawn a dashed horizontal line at  $170\degree$, which we use to approximately distinguish between ``bent'' and ``unbent'' DLRGs; note that due to the measurement uncertainties most objects with angles greater than $170\degree$ are consistent with having an angle of $180\degree$. 

In this plot, there are four DLRG-galaxy pairs, circled in red in Figure 2, whose velocity separations are all below 750 km/s and whose lobes subtend angles between $170\degree$ and $180\degree$.  The projected separation between the radio source and the central galaxy is also less than 0.2 Mpc for these four galaxies. We therefore hypothesize that they are the central galaxies of their respective systems. Central galaxies are not the focus of this paper. The evolution of their lobes can also be effected by buoyancy (\citealt{Gull1973}, \citealt{Churazov2000}, \citealt{Churazov2001}) as well as large-scale sloshing motions in the intracluster medium, especially if the cluster is not relaxed (there is some evidence that they are preferentially associated with merging clusters; \citealt{Sakelliou2000}). Due to the former issue, focus of this paper is on the behavior of satellite radio galaxies, and we neglect these four central galaxies in this work. We also assume the intracluster medium is quiescent; sloshing motions, if they exist, may introduce noise into our measurement. 

The dropoff with radius in projected density of DLRGs (discussed in the previous section) is also visible in Figure 2 (recall that the differential area increases linearly with projected radius). Based on our analysis in the previous section, the DLRG - galaxy matches with an impact parameter $\gapprox 2$ Mpc are consistent with being chance projections on the sky. Thus, while the three DLRG-galaxy pairs with $b = 3.5$ Mpc and $b = 6.8$ Mpc in Figure 2 (which are identified as \#25 and \#34 in Table 1) have radio lobes that show clear signs of bending, it is unlikely that the bending is caused by the galaxy group/cluster we have identified. As discussed in the previous section, we think that there are likely additional galaxy groups that lie closer to the DLRG but are below the detection threshold for their respective surveys. These two matches in particular lie at $z=0.437$ and $z=0.638$, which are near the upper ends of their respective redshift bins.

We can model the expected fraction of bent DLRGs using the chance projections on the sky, which we conservatively estimate from Figure 2 using the galaxies with impact parameter of at least 2 Mpc. There are 24 such galaxies, of which 7 are bent, corresponding to an expected bent fraction of 29\%. Excluding the four central galaxies in the red circle, the observed bent fraction within 1 (2) Mpc is 7/9 (9/16), corresponding to 78\% (56\%).  The seven bent galaxies within 1 Mpc are shown in Figure 3. Keeping the total number of galaxies within 1 (2) Mpc fixed, the expected number of bent galaxies within this impact parameter is 2.42 (4.31). Assuming binomial statistics, the probability of getting at least the observed number of bent galaxies, given the expected number, is $3.4\times10^{-4}$ ($5.6\times10^{-3}$). 

These probabilities indicate that the null hypothesis (DLRG bending being uncorrelated with the central galaxy) should be rejected at $3.6\sigma$  ($2.8\sigma$). We therefore conclude that the bending is correlated with the proximity of these DLRGs to the center of a nearby galaxy group or cluster.

\section{Discussion and Conclusions}

One of the results is that the density of DLRGs is declining more rapidly with radius than the density of galaxies in a typical cluster, which follows an NFW profile.  The projected density of DLRGs has a power-law slope in radius of $2.5^{+0.4}_{-0.3}$, or a space density decline of $r^{-m}$, where $m = 3.3-4.0$ which can be compared to the density of a NFW profile in the outer part of a cluster or group, where $m = 2.5-3$. There may be a few reasons for DLRGs to be more concentrated than the ensemble of galaxies.  One aspect is that the central dominant galaxy can be quite massive and the probability of it being a DLRG is enhanced relative to normal galaxies. 

Another factor is that the luminosity from the radio jet and radio lobes can be lower in the outer parts of the cluster because of the lower density.  Two important characteristic sizes of the radio structure scale as $n^{-1/2}$:  the recollimation of the jet \citep{alexander2006}; and the larger size when the lobes are in pressure balance with the surrounding medium \citep{kom1998}.  With these larger sizes, both the relativistic electron density and the magnetic field within the jets and lobes are likely lower, so the emissivity is less.  This is shown from simulations by \citet{hard2013,hard2014}, where the luminosity in lower density regions (due to steeper density laws for the ambient cluster medium) can be an order of magnitude less.  Lower luminosity lobes would be detected less frequently in flux-limited samples, so DLRGs at large radii from the center my exist but be undetected in the samples that we used.

Another aspect that we examined was the degree of bending as a function of distance from the cluster center.  Under the assumption that ram pressure is responsible, the ram pressure force is proportional to $n {\sigma}^2$, where $\sigma$ is the galaxy velocity dispersion of a cluster and $n$ is the ambient gas density.  The ambient density decreases rapidly, typically as $r^{-2}$ to $r^{-3}$ in a cluster (e.g \citealt{bahcall1994}), while the velocity dispersion has a very slow decline \citep{zhang11}.  The acceleration of the lobes is proportional to the ram pressure divided by the lobe mass, and if we assume that the lobe mass is independent of location in the cluster and that the lobe size is predicted to increase as $n^{-1/2}$ (then the area goes as $n^{-1}$), the acceleration is proportional to $n {\sigma}^2 \times n^{-1}/M_{lobe} \sim$ constant.  If the lifetime of DLRGs is independent of position in the cluster, the distance bent should be about the same, on average.  However, the DLRGs at large radii are expected to be longer, so as the bending angle is the displacement by ram pressure divided by the length of the DLRG, the ones furthest from the center should have smaller bending angles.

This expectation of smaller bending angles with distance is consistent with the data for $b \lapprox 2$ Mpc, but it is not proven by our data set.  Many more DLRGs would be needed to carry out this statistical test. As optical surveys become deeper, cluster and group catalogs will become much more complete. This should reduce the "background" of DLRGs whose associated cluster is not detected, enabling a much more precise measurement. This may help to constrain such parameters as the pressure in the lobes and the degree of density fluctuations in the intracluster medium, as well as definitively establishing the existence or non-existence of bent DLRGs outside of larger virialized halos. With a larger sample, it should be possible to study other interesting physics as well, such as the covering fraction of intercluster filaments beyond the virial radius.

\begin{figure}
  \centering
  \includegraphics[width=6.5 in]{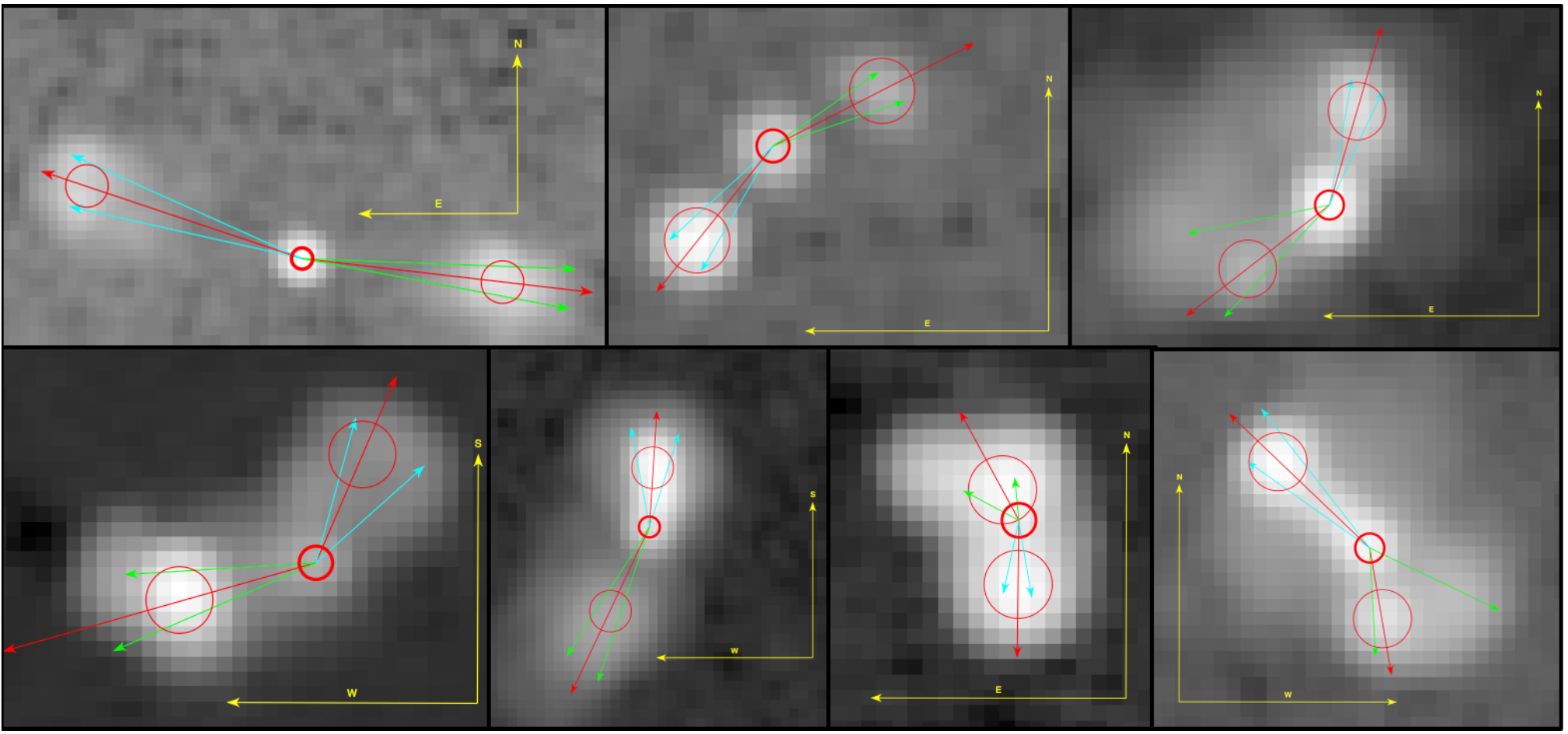}
  \caption{VLA FIRST images of the 7 bent DLRGs in the bottom left corner of Figure 2 - i.e. the seven DLRGs whose lobes subtend an angle less than $170\degree$ and who have an impact parameter less than 1 Mpc. Clockwise from top left, these objects have identification numbers 36, 19, 38, 41, 4, 40, and 37 in Table 1. In each image, the small, thick red circle is the location of the quasar at the DLRG core.  The larger, thinner red circles are the approximate locations of the DLRG lobes used to estimate the bending angle.  These positions were placed by hand based on the automatic estimated from the DBW catalog.  The red lines from the core through the center of the lobes were used to calculate the angle and the cyan and green lines to calculate the error on the angle.  The yellow lines are 30" and each image uses a logarithmic stretch to better show the faint structures.  }
  \label{fig:radio_images}
\end{figure}

\section{Acknowledgements}
We would like to thank Wim de Vries for sending us the list of DLRGs from their analysis, as well as Phillip Hughes and Eugene Churazov for helpful discussions and comments. We would like to acknowledge the Undergraduate Research Opportunities Program (UROP) at the University of Michigan as well. We thank the referee for a helpful report which improved the quality of the paper.

Funding for SDSS-III has been provided by the Alfred P. Sloan Foundation, the Participating Institutions, the National Science Foundation, and the U.S. Department of Energy Office of Science. The SDSS-III web site is http://www.sdss3.org/.

SDSS-III is managed by the Astrophysical Research Consortium for the Participating Institutions of the SDSS-III Collaboration including the University of Arizona, the Brazilian Participation Group, Brookhaven National Laboratory, Carnegie Mellon University, University of Florida, the French Participation Group, the German Participation Group, Harvard University, the Instituto de Astrofisica de Canarias, the Michigan State/Notre Dame/JINA Participation Group, Johns Hopkins University, Lawrence Berkeley National Laboratory, Max Planck Institute for Astrophysics, Max Planck Institute for Extraterrestrial Physics, New Mexico State University, New York University, Ohio State University, Pennsylvania State University, University of Portsmouth, Princeton University, the Spanish Participation Group, University of Tokyo, University of Utah, Vanderbilt University, University of Virginia, University of Washington, and Yale University.

\end{document}